\makeatletter{}

\documentclass[amsmath,amssymb,aps,reprint]{revtex4-1}

\makeatletter{}
\usepackage[T1]{fontenc}
\usepackage[mathletters]{ucs}
\usepackage[utf8]{inputenc}

\usepackage[hidelinks,colorlinks,breaklinks]{hyperref}

\usepackage{cleveref}

\crefname{equation}{equation}{equations}
\crefname{chapter}{chapter}{chapters}
\crefname{section}{section}{sections}
\crefname{appendix}{appendix}{appendices}
\crefname{enumi}{item}{items}
\crefname{footnote}{footnote}{footnotes}
\crefname{figure}{figure}{figures}
\crefname{table}{table}{tables}
\crefname{theorem}{theorem}{theorems}
\crefname{lemma}{lemma}{lemmas}
\crefname{corollary}{corollary}{corollaries}
\crefname{proposition}{proposition}{propositions}
\crefname{definition}{definition}{definitions}
\crefname{result}{result}{results}
\crefname{example}{example}{examples}
\crefname{remark}{remark}{remarks}
\crefname{note}{note}{notes}
 
\makeatletter{}
\usepackage{amsmath}
\usepackage{amssymb}
\usepackage{mathtools}
\usepackage{siunitx}
\usepackage{graphicx}

\usepackage{cool}

\usepackage[version=3]{mhchem}

\DeclarePairedDelimiter\abs{\lvert}{\rvert}
\DeclarePairedDelimiter\norm{\lVert}{\rVert}

\makeatletter
\let\oldabs\abs
\def\abs{\@ifstar{\oldabs}{\oldabs*}}
\let\oldnorm\norm
\def\norm{\@ifstar{\oldnorm}{\oldnorm*}}
\makeatother

\makeatletter{}\newcommand{\plotFitsInfo}{$W = \SI{2.2}{\micro \meter}$, $W_F = \SI{1.0}{\micro \meter}$, $σ_G = \SI{0.5}{\milli \siemens}$, $\rho_F = \SI{60}{\ohm \nano \meter}$, and $R_F = \SI{3.27}{\ohm}$ ($d = \SI{0.5}{\nano \meter}$ and $λ_F = \SI{0.06}{\micro \meter}$)} 
\makeatletter{} 

\makeatletter{}\usepackage{tikz}

\newcommand{\re}{\operatorname{Re}}

\newcommand{\rNL}{R_{\text{NL}}}
\newcommand{\rNLeff}{R_{\text{NL}}^{\text{SQ}}}
\newcommand{\sNL}{S_{\text{NL}}}
\newcommand{\rSQ}{R_{\text{SQ}}}

\definecolor{holo-blue}{HTML}{0099CC}
\definecolor{holo-purple}{HTML}{9933CC}
\definecolor{holo-green}{HTML}{669900}
\definecolor{holo-red}{HTML}{CC0000}

\hypersetup{
  linkcolor=holo-purple,
  citecolor=holo-green,
  filecolor=holo-red,
  urlcolor=holo-blue
}

\begin{document}
  \author{Evan Sosenko}
  \email{evan.sosenko@email.ucr.edu}
  \homepage{http://evansosenko.com/}

  \author{Huazhou Wei}
  \email{hwei002@ucr.edu}

  \author{Vivek Aji}
  \email{vivek.aji@ucr.edu}

  \affiliation{Department of Physics, University of California, Riverside, Riverside, California 92521, USA}

  \title{Effect of contacts on spin lifetime measurements in graphene}
  \date{\today}

  \makeatletter{}\begin{abstract}
  Injection, transmission, and detection of spins in a conducting channel
  are the basic ingredients of spintronic devices.
  Long spin lifetimes during transit are an important ingredient in realizing this technology.
  An attractive platform for this purpose is graphene, which has high mobilities
  and low spin-orbit coupling.
  Unfortunately, measured spin lifetimes are orders of magnitude smaller
  than theoretically expected.
  A source of spin loss is the resistance mismatch between
  the ferromagnetic electrodes and graphene.
  While this has been studied numerically,
  here we provide a closed form expression for Hanle spin precession
  which is the standard method of measuring spin lifetimes.
  This allows for a detailed characterization of the nonlocal spin valve device.
\end{abstract}
 
  \maketitle
  \makeatletter{}\section{Introduction}

Spintronic devices rely on the ability to inject, transport, manipulate, and detect spins
\cite{Wolf16112001, RevModPhys.76.323}.
The typical architecture involves ferromagnetic electrodes deposited on a conducting medium
\cite{1990ApPhL..56..665D, Jedema2001}.
Driving a current across the junction of a magnetic element and a nonmagnetic metal
leads to spin injection (also called spin accumulation)
\cite{PhysRevLett.55.1790, Jedema2001, Yang2008, PhysRevLett.94.196601}.
The injected spins either diffuse in nonlocal spin valve geometry,
or are driven by applied fields across the conducting channel.
The former has the advantage that the observed spin signal
is not corrupted by accompanying charge current.
During this transit, scattering processes dephase the spins
and thus degrade the chemical potential imbalance between spins of opposite orientation.
The residual difference is detected by a ferromagnetic electrode
whose magnetization can be flipped by applying external fields.

The performance of devices is determined by a number of parameters
associated with the basic processes described above.
The efficiency of spin injection, the diffusion length
(or equivalently the diffusion constant and spin relaxation time),
the distance between the injector and detector,
and resistivities of various components such as the electrodes,
the junction, and the conducting channel,
are some of the ingredients that contribute to the measured magnetoresistance.
As such, having good injection efficiency coupled with long spin lifetimes
is crucial for the viability of spintronic applications.
The discovery of graphene
\cite{Novoselov22102004}
has been of particular interest in this regard
because of its tunable conductivity, high mobility, and low spin-orbit coupling.
Moreover, the two dimensional nature allows for efficient device design and spin manipulation.
Theoretical estimates for spin lifetimes of a few microseconds
\cite{PhysRevB.74.155426, Trauzettel2007}
are leading to a concerted effort in realizing spin based transistors and spin valves
\cite{Tombros2007, JJAP.46.L605, Cho2007, PhysRevLett.101.046601, 1704408, Han2012, Han2012369, PhysRevB.80.241403}.

Unfortunately, the best measured spin lifetimes
via the Hanle spin precession technique are in the
\SIrange{50}{200}{\pico \second} range
\cite{PhysRevB.80.241403, Tombros2007, PhysRevB.80.214427, PhysRevLett.104.187201}.
The large discrepancy is yet to be explained.
The linear scaling of spin and transport lifetimes
\cite{PhysRevB.80.241403}
suggested that the dominant scattering mechanism in the conducting channeling
is of the Elliot-Yafet
\cite{PhysRev.96.266}
type.
Surprisingly, in the regime of small spin lifetimes
($∼ \SI{100}{\pico \second}$),
Coulomb scattering was shown not to be the dominant mechanism
\cite{PhysRevLett.104.187201}.
The more important determining factor of the lifetime
was found to be the nature of the interface between
the magnetic electrode and the conducting channel.
Tunneling contacts suppress spin relaxation, and lifetimes of \SI{771}{\pico \second}
were reported at room temperature, increasing to
\SI{1.2}{\nano \second} at \SI{4}{\kelvin}
\cite{PhysRevLett.107.047207}.
On the other hand, low resistance barriers lead to considerable
uncertainty in the determination of the lifetimes.

Over the last few years, characterizing the nature of the spin dynamics
at the interface has garnered much attention.
A key contribution in this effort is the generalization of the standard theoretical approach
of calculating the nonlocal magnetoresistance with and without the magnetic field.
Recent efforts study the effect of including the contact resistance
\cite{PhysRevB.80.214427, PhysRevB.67.052409},
and alternatively relaxing the normally infinite boundary conditions
in favor of a finite channel size
\cite{1404.6276v1}.
The approach relies on numerically solving the Bloch equation
to generate Hanle precession curves and then fitting observed data.

In this paper, we present the closed form expression
for the precession curves with finite contact resistance,
and analytically discuss the various parameters regimes
that show qualitatively different behaviors.
The fits to data reproduce the results in the literature
and provide a means to understand the effect of the contacts
which were previously obtained by numerical simulations.

The paper is organized as follows.
In \cref{s:model} we provide the basic model, define the relevant parameters,
and present an expression for the nonlocal resistance $\rNL$.
The primary result is given by \cref{eq:f}.
In \cref{s:fits} the solution for $\rNL$ is fitted to data.
In \cref{s:regimes} we analyze the various regimes which are determined by
the diffusion length, length of the device, and the contact resistance.
\Cref{s:summary} ends with a summary of the results and future directions.
 
  \makeatletter{}\section{Model}
\label{s:model}

\begin{figure}
  \caption{
    The geometry of the nonlocal spin valve analyzed in this paper is shown.
    There are two ferromagnetic electrodes placed on a conducting channel.
    Current $I$ flows into the left electrode,
    while the potential $V$ is measured at the right electrode.
    The nonlocal resistance is defined as the ratio $V / I$.
    For spin dependent phenomena, the relevant quantity of interest
    is the difference between the nonlocal resistance for the parallel
    and antiparallel orientations of magnetization of the two electrodes.
  }
  \label{fig:nonlocal_spin_valve}
  \makeatletter{}
\usetikzlibrary{arrows}
\usetikzlibrary{decorations.markings}

\pgfdeclarelayer{semiconductor}
\pgfdeclarelayer{contacts}
\pgfsetlayers{semiconductor,contacts,main}

\tikzstyle{semiconductor}=[fill=gray]
\tikzstyle{contact}=[fill=lightgray,opacity=0.5]
 
  \begin{tikzpicture}[scale=0.7]
    \makeatletter{}

\begin{pgfonlayer}{contacts}
  \path[contact] (1,-1) rectangle ++(2,5);
  \path[contact] (5,-1) rectangle ++(2,5);

  \node at (2,3.5) {$F$};
  \node at (6,3.5) {$F$};

  \draw[serif cm-serif cm] (1,-1.1) -- node[below] {$W_F$} ++(2,0);

  \begin{scope}[very thick,
    decoration={markings,
      mark=at position 0.7 with {\arrow{stealth}}}]
    \draw[o-,postaction={decorate}] (2,5) -- ++(0,-1);
  \end{scope}

  \begin{scope}[very thick,
    decoration={markings,
      mark=at position 0.4 with {\arrow{stealth}}}]
    \draw[-o,postaction={decorate}] (0,1.5) -- ++(-1,0);
  \end{scope}

  \draw[shift={(2,4.4)}] node[right] {$I$};
  \draw[shift={(-0.3,1.5)}] node[above] {$I$};
\end{pgfonlayer}

\begin{pgfonlayer}{semiconductor}
  \draw[semiconductor] (0,0) rectangle ++(8,3);

  \node at (7.5,12pt) {$N$};

  \draw[serif cm-serif cm] (8.1,0) -- node[right] {$W$} ++(0,3);
\end{pgfonlayer}

\draw[very thick] (6,4) -- ++(0,0.5) -- ++(2,0)
  ++(0.4,0) circle [radius=0.4] node {$V$}
  ++(0,-0.4) -- ++(0,-1.6) -- ++(-0.4,0);

\draw[shift={(2,0)}] (0pt,2pt) -- (0pt,-2pt) node[below] {$x = 0$};
\draw[shift={(6,0)}] (0pt,2pt) -- (0pt,-2pt) node[below] {$x = L$};

\fill (0,0) circle [radius=1pt];
\draw (0,0) circle [radius=3pt] node[left] {$z, -z'$};

\begin{scope}[decoration={
  markings,
  mark=at position 1 with {\arrow{stealth}}}]
  \draw[postaction={decorate}] (0,0) -- (0,3.5) node[left] {$y$};
\end{scope}
 
  \end{tikzpicture}
\end{figure}
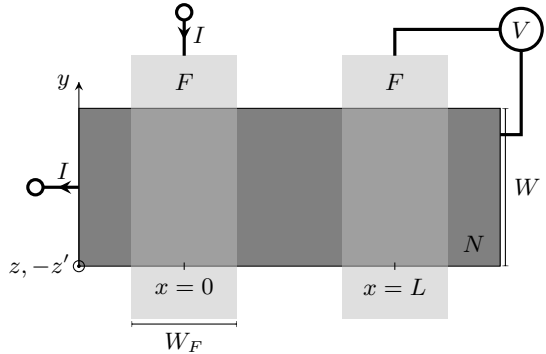

The assumed device geometry is shown in \cref{fig:nonlocal_spin_valve}.
Two ferromagnetic contacts ($F$) are deposited on the normal semiconductor ($N$).
A spin-polarized current $I$ is injected through the contact at $x = 0$
and flows in the $x ≤ 0$ region of the semiconductor.
The voltage difference $V$ is measured at $x = L$ between the contact and the semiconductor.
The nonlocal resistance is $\rNL = V / I$ \cite{PhysRevB.67.052409}.

Spin transport is modeled by identifying two spin channels
and their associated three-component spin electrochemical potentials $μ_{↑↓}$.
The majority channel is labeled as up, while the minority channel is labeled as down.
The voltage difference is proportional to the spin accumulation
$μ_s = \left( μ_↑ - μ_↓ \right) / 2$ at $x = L$.
The spin accumulation in the semiconductor
is assumed to satisfy the steady-state Bloch diffusion equation
\begin{equation}
  D ∇^2 μ_s^N - \frac{μ_s^N}{τ} + ω × μ_s^N = 0 .
\end{equation}
The key parameters are
the contact spacing $L$,
the diffusion constant $D$,
the spin lifetime $τ$,
the spin diffusion length $λ = \sqrt{D τ}$,
and $ω = \left( g μ_B / ℏ \right) B$ which is proportional to
the applied magnetic field $B$ and the gyromagnetic ratio $g = 2$.

For contacts which cover the width of the channel,
the transport is uniform along $y$.
Since the channel is two-dimensional, $μ_s^N$ will only vary along $x$.
We enforce the boundary condition $μ_s^N → 0$ at $x → {±} ∞$
and the continuity of the current and spin current.
A detailed derivation is given in \cref{s:appendix} and reveals
\begin{equation}
  \label{eq:nonlocal_resistance}
  \rNL^{±} = {±} p_1 p_2 R_N f .
\end{equation}
The overall sign corresponds to parallel and antiparallel ferromagnetic alignments.
Specifically, we find a resistance scale
\begin{equation}
  R_N = \frac{λ}{W L} \frac{1}{σ^N} ,
\end{equation}
and the function
\begin{multline}
  \label{eq:f}
  f = \re \left( \left\{
        \vphantom{
          \frac{
            \sinh{ \left[ \left( L / λ \right) \sqrt{1 + i ω τ} \right] }
          }{\sqrt{1 + i ω τ}}
        }
        2 \left[ \sqrt{1 + i ω τ} + \frac{λ}{2} \left( \frac{1}{r_0} + \frac{1}{r_L} \right) \right]
        e^{\left( L / λ \right) \sqrt{1 + i ω τ}}
        \right. \right. \\ \left. \left.
        + \frac{λ^2}{r_0 r_L} \frac{
            \sinh{ \left[ \left( L / λ \right) \sqrt{1 + i ω τ} \right] }
          }{\sqrt{1 + i ω τ}}
      \right\}^{-1} \right) .
\end{multline}

Note that $f$ is unitless and depends only on the scales $L / λ$, $ω τ$, and $λ / r_i$.
The parameters $r_i$ with $i$ either $0$ for the left contact or $L$ for the right are
\begin{equation}
  \label{eq:r-parameter}
  r_i = \frac{R_F + R_C^i}{\rSQ} W ,
\end{equation}
where $R_F$ is the resistance of the ferromagnet
and $R_C^i$ are the individual contact resistances,
$W$ is the graphene flake width, and
\begin{equation}
  \label{eq:square_resistance}
  \rSQ = W / σ^N ,
\end{equation}
is the graphene square (sheet) resistance
given in terms of the semiconductor conductivity $σ^N$.
The resistances $R_F$ and $R_C^i$ are the effective resistances
of a unit cross sectional area.
They are defined in \cref{eq:contact.resistance,eq:ferromagnet.resistance}.
To obtain an expression in terms of the ohmic resistances,
one must make the substitutions
$R_F → W_F W R_F$ and $R_C^i → W_F W R_C^i$,
where $W_F$ is the contact width, i.e., $W_F W$ is the contact area.
We will use the same symbols for either resistance type when the meaning is clear.
The polarizations $p_1$ and $p_2$, defined in \cref{eq:polarizations},
model the effective current injection.
They depend on the resistances and the spin polarizations
of the semiconductor and the individual contacts.

The expression $Δ \rNL = \abs{\rNL^+ - \rNL^-}$
measures the difference in signal between parallel and antiparallel field alignments.
We combine $P^2 = \abs{p_1 p_2}$
\footnote{
  Assuming the polarizations $P_σ^F$ and $P_Σ^L$ have the same sign bounds $P ≤ 1$.
},
and write
\begin{equation}
  \label{eq:nonlocal_resistance.difference}
  Δ \rNL = 2 P^2 R_N \abs{f} ,
\end{equation}
with
\begin{equation}
  R_N = \frac{λ}{W} \frac{1}{σ_G} ,
\end{equation}
where $σ_G = σ^N L$ is the graphene conductance normally given in units of
$\si{\milli \siemens} = \left( \si{\milli \ohm} \right)^{-1}$.
 
  \makeatletter{}\section{Fits}
\label{s:fits}

Data presented in figure 4 from \cite{PhysRevLett.105.167202}
was fit to the model presented here.
Fits done using Python and matplotlib \cite{Hunter:2007}.
Links to the source code along with instructions
on how to create similar fits and figures are available online
\footnote{
  An online portal with links to the code used to prepare this work is located at
  \href{http://evansosenko.com/spin-lifetime/}{evansosenko.com/spin-lifetime/}.
}.

We assume similar contacts, $R_C = R_C^0 = R_C^L$.
The resistance of the ferromagnet \ce{Co} is computed as
$R_F = \rho_F λ_F / A_J$,
where $\rho_F$ is the \ce{Co} resistivity,
$λ_F$ is the spin diffusion length of \ce{Co},
and $A_J$ is the junction area estimated at $A_J = W d$,
with $d$ between \SIrange[range-phrase={ and }]{0.5}{50}{\nano \meter}
\cite{PhysRevLett.105.167202}.
Hanle fits were done using a simple least squares algorithm
with nonnegative parameters $τ$, $D$, $R_C$, and $P$.
The polarization $P$ was constrained between zero and one.

\begin{figure}
  \caption{
    Data in figure 4 from \cite{PhysRevLett.105.167202}
    fit to \cref{eq:nonlocal_resistance.difference}
    or \cref{eq:nonlocal_resistance}
    with the following values: \plotFitsInfo.
    The contact type (tunneling, pinhole, or transparent)
    and the contact separation $L$ varies.
  }
  \label{fig:nonlocal_resistance.fits}
  \includegraphics[width=\columnwidth]{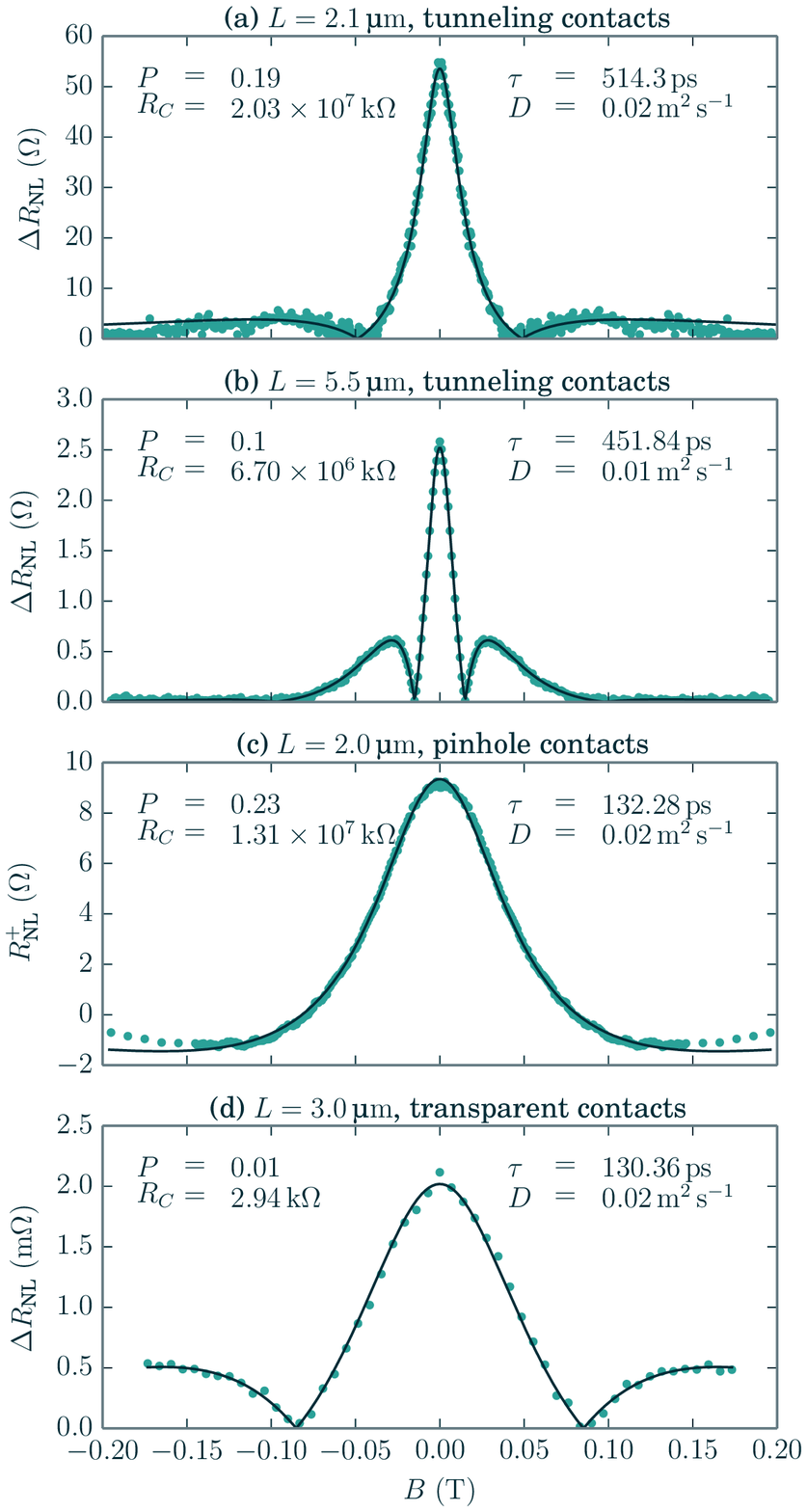}
\end{figure}

\Cref{fig:nonlocal_resistance.fits} shows fits of
$Δ \rNL$ given by \cref{eq:nonlocal_resistance.difference}
for devices with tunneling and transparent contacts,
and $\rNL^+$ given by \cref{eq:nonlocal_resistance}
for a device with pinhole contacts
\footnote{
  Parallel and antiparallel data for this device was only available
  at dissimilar field values, thus $Δ \rNL$ could not be fit.
}.
Fits (a), (b), and (c) with tunneling and pinhole contacts give large
$R_C ∼ \SI{e7}{\kilo \ohm}$ and lifetimes equivalent to fitting with $R_C → ∞$,
while (d) with transparent contacts gives a reduced $R_C ∼ \SI{3}{\kilo \ohm}$
and a lifetime increased by at most a factor of two
(compare to \SI{78}{\pico \second} for $R_C → ∞$).
For tunneling contacts, the polarization $P$ is \SIrange{25}{60}{\percent}
smaller than the lower bound given in \cite{PhysRevLett.105.167202},
while for transparent contacts, $P$ is reduced by an order of magnitude.

Note that we have used $R_C$ as a fitting parameter.
In most devices, this quantity can be experimentally determined,
thus further constraining the fitting algorithm.
As we will discuss further in the next section, a fact that becomes apparent
from our analytic result is that the relevant scale is $λ / r$.
Once $r$ becomes larger than $λ$,
all of the corrections to the $R_C → ∞$ limit Hanle curves become very small.
In other words, once $r ≫ λ$,
the fit is insensitive to the actual value of the contact resistance.
The fact that we quote a resistance of order \SI{e7}{\kilo \ohm} in fits (a), (b), and (c)
in \cref{fig:nonlocal_resistance.fits}
results from the built-in accuracy we demand of the fitting algorithm.
A good fit can be obtained for any $r$ as long as it is larger than $λ$.
 
  \makeatletter{}\section{Regimes}
\label{s:regimes}

In this section we discuss the various limits of
the expression describing the Hanle precession curve.
First, we show that the commonly used results for zero magnetic field
and tunneling contacts are correctly reproduced.
Next, we discuss regimes where appropriate scaling will give non-unique Hanle fits.
In the following, we consider the case $r = r_0 = r_L$ of similar contacts.

In the limit of tunneling contacts, $R_C^0, R_C^L ≫ R_F$.
Putting $r_0, r_L → ∞$ gives $p_1 p_2 → \left( P_Σ^L \right)^2$ and
\begin{equation}
  f^∞ = \re{\frac{e^{- \left( L / λ \right) \sqrt{1 + i ω τ}}}{2 \sqrt{1 + i ω τ}}} ,
\end{equation}
which is of the same form as found in appendix B of
\cite{PhysRevB.37.5312}
(we will denote this limit with the superscript $∞$).
Fitting with this expression was found to give results equivalent
to fitting with the Hanle equation
\begin{equation}
  \label{eq:hanle_integral}
  \rNL^{±} = {±} \sNL ∫_0^∞ \frac{e^{-t / τ}}{\sqrt{4 π D t}}
             \exp{\left[- \frac{L^2}{4 D t} \right]} \cos{ω t} \: dt .
\end{equation}
The agreement is expected as an explicit integration of \cref{eq:hanle_integral}
yields the same analytic expression with the identification
$\sNL = {p_1 p_2 D} / {W σ_G}$.
In the additional limit of zero magnetic field,
\begin{equation}
  Δ \rNL = \left( P_Σ^L \right)^2 R_N e^{- L / λ} ,
\end{equation}
which agrees with equation (6) in
\cite{PhysRevB.67.052409}.

Let $f_0$ denote $f$ at zero magnetic field,
\begin{equation}
  f_0 = \left[ 2 \left( 1 + λ / r \right) e^{L / λ} + \left( λ / r \right)^2 \sinh{L / λ} \right]^{-1} ,
\end{equation}
which agrees with equation (3) in
\cite{PhysRevB.80.214427}.

To further explore the nature of the Hanle curves,
we exploit the fact that it only depends on
the dimensionless ratios $λ / r$, $L / λ$, and $ω τ$.
The only other parameter of the conducting channel that enters the expression
is the overall scale $λ$ in $R_N$.
The expression $f$ contains three terms
which are of zeroth, first, and second order in $λ / r$.
Thus, as the contact resistance decreases,
one goes from a device dominated by the first term to one dominated by the last.
But precisely how this comes about depends on the value of $ω τ$.

For infinite contact resistance, it was pointed out that any rescaling
of $g$, $τ$ and $D$ that leaves $λ$ and $ω τ$ unchanged
leads to the same Hanle precession curves
\cite{Swartz2013}.
Our result shows that the same is also true
when the contact resistance is taken into account.
In numerical simulations, interesting features were observed
when $L / λ ≪ 1$ and $r / λ ≪ 1$
\cite{PhysRevB.86.235408}.

To compare across regimes, we first normalize the data to its value at zero magnetic field.
In devices where $λ / r ≫ 1$, the normalization factor is
\begin{equation}
  f_0 = \frac{2 e^{- L / λ}}{\left( λ / r \right)^2} .
\end{equation}
In this regime, if $D$ is not very different from the infinite contact resistance value,
then the lifetime can be large, i.e., $τ ≫ \SI{1}{\nano \second}$.
As one tunes the magnetic field $\sqrt{ω τ} ≫ 1$, for small values of the field,
and for much of the curve, we can approximate $1 + i ω τ ≈ i ω τ$.
An interesting consequence of this is that the zero of the Hanle precession curve
becomes independent of the scattering time.
Note that the product
\begin{equation}
   \frac{L}{λ} \sqrt{ω τ} = L \sqrt{\frac{D}{ω}} ,
\end{equation}
which appears in the exponential and oscillating factors below,
is independent of the lifetime.
As one further tunes the magnetic field, the Hanle curve is given by
\begin{equation}
  \label{eq:regime.1.f}
  f = \frac{\sqrt{ω τ}}{\left( λ / r \right)^2}
      e^{- \left( L / λ \right) \sqrt{ω τ / 2}}
      \sin{\left[ \frac{L}{λ} \sqrt{\frac{ω τ}{2}} + \frac{π}{4} \right]} ,
\end{equation}
as long as $λ / r ≫ \sqrt{ω τ} ≫ 1$.
In this limit, the nonlocal resistance scales as
\begin{equation}
  \label{eq:regime.1.nonlocal_resistance}
  Δ \rNL ∝ \frac{λ \sqrt{ω τ}}{\left( λ / r \right)^2} = r^2 \sqrt{\frac{ω}{D}} ,
\end{equation}
and the normalized nonlocal resistance as
\begin{equation}
  f / f_0 ∝ \sqrt{ω τ} .
\end{equation}

On further increasing the field,
$\sqrt{ω τ} ≫ λ / r ≫ 1$, we get
\begin{equation}
  \label{eq:regime.2.f}
  f = \frac{1}{2 \sqrt{ω τ}}
      e^{- \left( L / λ \right) \sqrt{ω τ / 2}}
      \cos{\left[ \frac{L}{λ} \sqrt{\frac{ω τ}{2}} + \frac{π}{4} \right]} .
\end{equation}
In this limit, the nonlocal resistance scales as
\begin{equation}
  \label{eq:regime.2.nonlocal_resistance}
  Δ \rNL ∝ \frac{λ}{\sqrt{ω τ}} = \sqrt{\frac{D}{ω}} ,
\end{equation}
and the normalized nonlocal resistance as
\begin{equation}
  \label{eq:regime.2.ratio}
  f / f_0 ∝ \frac{\left( λ / r \right)^2}{\sqrt{ω τ}} = D \sqrt{\frac{τ}{ω r^4}} .
\end{equation}

In the limits of \cref{eq:regime.1.f,eq:regime.2.f},
the zeros of the Hanle fit are independent of the lifetime
and are determined by $D$ though the condition
\begin{equation}
  L \sqrt{\frac{D}{2 ω}} + \frac{π}{4} = \frac{n π}{2} ,
\end{equation}
where $n = 0$ for \cref{eq:regime.1.f} and $n = 1$ for \cref{eq:regime.2.f}.

\begin{figure}[!b]
  \caption{
    Data in figure 4 (d) from \cite{PhysRevLett.105.167202}
    fit to \cref{eq:nonlocal_resistance}
    with the same values as in
    \cref{fig:nonlocal_resistance.fits} (d).
    Fits with lifetimes that differ by four orders of magnitude
    were obtained by using different starting values for $τ$.
    These fits are otherwise similar with the exception of the lifetime,
    demonstrating the $τ$-independent scaling in \cref{eq:regime.1.nonlocal_resistance}.
    The $χ^2$ for \cref{fig:nonlocal_resistance.fits} (d)
    is \SI{2}{\percent} less than the $χ^2$ for
    \cref{fig:nonlocal_resistance.large_lifetime}.
  }
  \label{fig:nonlocal_resistance.large_lifetime}
  \includegraphics[width=\columnwidth]{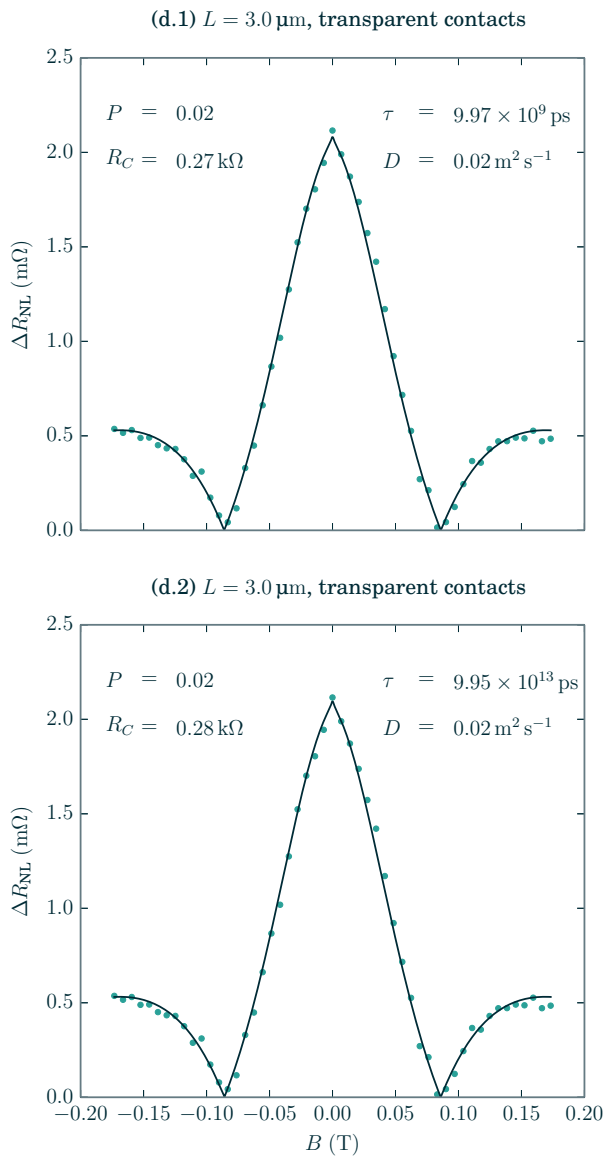}
\end{figure}

Note that fitting is insensitive to $τ$ in the limit of
\cref{eq:regime.1.nonlocal_resistance}
or \cref{eq:regime.2.nonlocal_resistance}.
As an example of this,
\cref{fig:nonlocal_resistance.large_lifetime} shows nearly identical fits
with lifetimes that differ by four orders of magnitude.
These fits were obtained by choosing large starting values for $τ$.
For \cref{fig:nonlocal_resistance.fits} (d)
and \cref{fig:nonlocal_resistance.large_lifetime},
$χ^2 ∼ \SI{7e-8}{}$, but the $χ^2$ for
\cref{fig:nonlocal_resistance.fits} (d)
is \SI{2}{\percent} less than the $χ^2$ for
\cref{fig:nonlocal_resistance.large_lifetime}.
In \cref{fig:nonlocal_resistance.fits} (d),
$λ / r ≫ \sqrt{ω τ}$ and $ω τ ∼ 1$ for most of the curve,
so the approximation $1 + i ω τ ≈ i ω τ$ does not hold.
However, \cref{fig:nonlocal_resistance.large_lifetime}
is in the limit of \cref{eq:regime.1.nonlocal_resistance}
for all points (save the origin).
Thus, in limit of small $r$, the fitted value of $τ$ is unreliable
unless one carefully controls the fitting procedure.

The evolution of the expression for the Hanle curve
is an interesting insight into the behavior of the device.
Fitting data on devices with small contact resistances
with the functional form applicable to infinite contact resistance
yields unreliable parameters.
In particular, they were numerically shown to severely underestimate the spin lifetime
\cite{PhysRevB.86.235408}.

Further analytic progress can be made if one assumes that
lifetimes as estimated with infinite contact resistance are long
enough that the approximation of $\sqrt{ω τ} ≫ λ / r ≫ 1$
is still valid for much of the data being analyzed.
For this case, at infinite contact resistance,
the normalized nonlocal resistance is given by
\begin{equation}
  \frac{f^∞}{f^∞_0} = \frac{1}{ \sqrt{ω τ}}
                      e^{- \left( L / λ \right) \sqrt{ω τ / 2}}
                      \cos{\left[ \frac{L}{λ} \sqrt{\frac{ω τ}{2}} + \frac{π}{4} \right]} .
\end{equation}
Provided $D$ remains constant,
this will yield the same curve with finite contact resistance if
\begin{equation}
  \label{eq:lifetime_scale.infinity}
  \frac{1}{τ^∞} = D^2 \frac{τ}{r^4} .
\end{equation}
In other words, if we fix $τ$ and ask what happens to the fitted value
assuming infinite contact resistance as a function of decreasing $r$,
\cref{eq:lifetime_scale.infinity} shows that it will decrease as well.
For $D$ fixed, $τ^∞ ∝ r^4$.
While the general trend is consistent with
\cite{PhysRevB.86.235408},
the quantitative agreement is limited by
the approximations made for analytic convenience.
 
  \makeatletter{}\section{Summary}
\label{s:summary}

In this paper we have analyzed the effect of contact resistance
on spin lifetimes determined via the Hanle spin prescession technique in nonlocal spin valves.
The general expression for the precession curves given in \cref{eq:f} is the main new result.
While aspects of the discussed phenomena have been addressed numerically before,
an analytic solution is obtained here which allows for detailed characterization of the device.
In particular, general features of scaling as well as various limits and regimes can be analyzed.
In addition, the solution allows for fitting data using standard curve fitting algorithms.

\begin{acknowledgments}
  We acknowledge useful discussions with Roland Kawakami, Adrian Swartz, and Sung-Po Chao.
  The work was partially supported by a UCR Senate Research Grant.
\end{acknowledgments}
 
  \makeatletter{}\appendix
\section{Computation}
\label{s:appendix}

In this appendix we derive an expression for
the nonlocal resistance for finite contact resistance.
We first present the key definitions and critical boundary conditions.
We then derive the relation between the nonlocal resistance
and the spin chemical potential at the far contact, $μ_s^N (L)$.
Finally, we solve the diffusion equation inside the semiconductor to find $μ_s^N (L)$.

\subsection{Definitions}

Many of the definitions and results in this section are taken from
\cite{ActaPhysicaSlovaca.57.4_5.565-907}.
The chemical potential and spin chemical potential are defined in terms
of the spin-up and spin-down chemical potentials,
\begin{subequations}
  \label{eq:potentials}
  \begin{alignat}{2}
    & μ   && = \frac{1}{2} \left( μ_↑ + μ_↓ \right) , \\
    & μ_s && = \frac{1}{2} \left( μ_↑ - μ_↓ \right) .
  \end{alignat}
\end{subequations}
The material conductances and polarization are defined in terms
of the spin-up and spin-down conductances,
\begin{subequations}
  \label{eq:conductances}
  \begin{alignat}{2}
    & σ   && = σ_↑ + σ_↓ , \\
    & σ_s && = σ_↑ - σ_↓ , \\
    \label{eq:material.polarization}
    & P_σ && = \frac{σ_s}{σ} .
  \end{alignat}
\end{subequations}
The gradient of the chemical potentials drives a current and spin current,
\begin{subequations}
  \label{eq:currents}
  \begin{alignat}{3}
    & J_{↑↓} && = σ_{↑↓} ∇μ_{↑↓} , \\
    \label{eq:currents.current}
    & J      && = J_↑ + J_↓ & = σ   ∇μ + σ_s ∇μ_s , \\
    \label{eq:currents.spincurrent}
    & J_s    && = J_↑ - J_↓ & = σ_s ∇μ + σ   ∇μ_s .
  \end{alignat}
\end{subequations}
To indicate the material, any of the above can have a superscript
$N$ (normal semiconductor) or $F$ (ferromagnet).

The contact conductances and polarization are defined in terms
of the spin-up and spin-down contact conductances,
\begin{subequations}
  \label{eq:contact_conductances}
  \begin{alignat}{2}
    & Σ   && = Σ_↑ + Σ_↓ , \\
    & Σ_s && = Σ_↑ - Σ_↓ , \\
    \label{eq:contact.polarization}
    & P_Σ && = \frac{Σ_s}{Σ} .
  \end{alignat}
\end{subequations}
The mismatch of the chemical potentials across the contact
drives a current and spin current,
\begin{subequations}
  \label{eq:contact_currents}
  \begin{alignat}{2}
    & J_{↑↓}^C && = Σ_{↑↓} \left( μ^N_{↑↓} - μ^F_{↑↓} \right)_c , \\
    & J^C      && = J_↑^C + J_↓^C , \\
    & J_s^C    && = J_↑^C - J_↓^C .
  \end{alignat}
\end{subequations}
The subscript $c$ will always denote the function evaluated at the contact.

We will use the term current to refer to $J$,
when in fact this is a particle current density.
For constant $J$, the physical charge current $I$ will be related to $J$
by a relation $I = - A J / e$ for some characteristic area $A$.

To reduce the number of subscripts and superscripts in the following,
we adopt the notation for the potentials
\begin{subequations}
  \begin{equation}
    \begin{aligned}
    & \begin{alignedat}{2}
        & u && = μ^N_s , \\
        & v && = μ^N   ,
      \end{alignedat}
    & \begin{alignedat}{2}
        & φ && = μ^F_s , \\
        & ψ && = μ^F   ,
      \end{alignedat}
    \end{aligned}
  \end{equation}
  and currents
  \begin{equation}
    \begin{alignedat}{2}
      & \jmath   && = J_s   , \\
      & J_c && = J^C   , \\
      & \jmath_c && = J_s^C .
    \end{alignedat}
  \end{equation}
\end{subequations}

We rewrite \cref{eq:contact_currents} as
\begin{subequations}
  \label{eq:contact_currents.2}
  \begin{alignat}{3}
    \label{eq:contact_currents.2.current}
    & J_c && = Σ   \left( v_c - ψ_c \right) && + Σ_s \left( u_c - φ_c \right) , \\
    \label{eq:contact_currents.2.spincurrent}
    & \jmath_c && = Σ_s \left( v_c - ψ_c \right) && + Σ   \left( u_c - φ_c \right) ,
  \end{alignat}
\end{subequations}
and \cref{eq:conductances,eq:currents} as
\begin{equation}
  \label{eq:bdry_current}
  \jmath = P_σ J + 4 \frac{σ_↑ σ_↓}{σ} ∇μ_s .
\end{equation}

Using \cref{eq:contact_conductances,eq:contact_currents.2},
\begin{equation}
  \label{eq:bdry_current_contact}
  \jmath_c = P_Σ^i J_c + {R_C^i}^{-1} \left( u_c - φ_c \right) ,
\end{equation}
where the contact resistance is
\begin{equation}
  \label{eq:contact.resistance}
  R_C^i = \frac{Σ^i}{4 Σ_↑^i Σ_↓^i} .
\end{equation}
The superscript $i$ allows for contacts with difference conductances.

\subsection{Boundary conditions}

In this sections, we derive the relations between the potentials and the currents
This corresponds to the needed boundary conditions.

\subsubsection{Semiconductor}

For the semiconductor, $σ^N_↑ = σ^N_↓ = σ^N / 2$, so $P_σ^N = 0$.
Evaluating \cref{eq:bdry_current} at the contact gives
\begin{equation}
  \label{eq:bdry_current.semiconductor}
  \jmath^N_c = σ^N ( ∇u )_c .
\end{equation}

\subsubsection{Ferromagnet}

For the ferromagnet, one assumes $μ_s^F$ satisfies
the one dimensional diffusion equation.
We choose the $z'$ coordinate antiparallel to $z$ with origin at the contact.
The equation
\begin{equation}
  \label{eq:diffusion.ferromagnet}
  φ'' \left( z' \right) - k_F^2 φ \left( z' \right) = 0 ,
\end{equation}
with the boundary condition
$\lim_{z' → - ∞} φ(z') = 0$
has solution
\begin{equation}
  \label{eq:diffusion.ferromagnet.solution}
  φ(z') = φ_c e^{k_F z'} ,
\end{equation}
where $φ_c = φ(0)$ is a yet undetermined constant.
Putting this into \cref{eq:bdry_current} and evaluating it at the contact gives
\begin{equation}
  \label{eq:bdry_current.ferromagnet}
  \jmath^F_c = P_σ^F J^F_c + R_F^{-1} φ_c ,
\end{equation}
where the ferromagnet resistance is
\begin{equation}
  \label{eq:ferromagnet.resistance}
  R_F = \frac{σ^F}{ 4 σ_↑^F σ_↓^F k_F } .
\end{equation}
Here, $λ_F = 1 / k_F$ is the spin diffusion length in the ferromagnet.

\subsubsection{Continuity assumptions}

At the contact, the current and spin current are assumed continuous,
\begin{subequations}
  \label{eq:continuity.current}
  \begin{alignat}{3}
    & J_c && = J^F_c && = J^N_c , \\
    & \jmath_c && = \jmath^F_c && = \jmath^N_c .
  \end{alignat}
\end{subequations}
Using \cref{eq:bdry_current.ferromagnet,eq:bdry_current_contact,eq:continuity.current}
we find the relation
\begin{subequations}
  \label{eq:bdry_solutions}
  \begin{equation}
    \label{eq:bdry_solutions.current}
    \left( P_σ^F R_F + P_Σ^i R_C^i \right) J_c = \left( R_F + R_C^i \right) \jmath_c - u_c ,
  \end{equation}
  and that $φ_c$ is determined by
  \begin{equation}
    \label{eq:bdry_solutions.potential}
    R_F^{-1} φ_c = \frac{ \left( P_Σ^i - P_σ^F \right) R_C^i \jmath_c + P_σ^F u_c}{P_σ^F R_F + P_Σ^i R_C^i} .
  \end{equation}
\end{subequations}
In the special case of zero current at the contact ($J_c = 0$),
\cref{eq:bdry_solutions} reduces to
\begin{subequations}
  \label{eq:bdry_solutions.zero}
  \begin{alignat}{2}
    \label{eq:bdry_solutions.zero.current}
    & \jmath_c && = \frac{1}{R_F + R_C^i} u_c   , \\
    \label{eq:bdry_solutions.zero.potential}
    & φ_c && = \frac{R_F}{R_F + R_C^i} u_c .
  \end{alignat}
\end{subequations}

\subsection{Nonlocal resistance}
\label{s:appendix:nonlocal_resistance}

In this section we derive the precise relation between $\rNL$ and $μ_s^N (L)$.
Note that we may write in general, for some $\bar{μ}$,
\begin{equation}
  μ = \bar{μ} + P_σ μ_s ,
\end{equation}
and, following
\cite{PhysRevB.67.052409},
define the voltage due to the difference in the chemical potentials across the contacts by
\begin{equation}
  V_c = \left( \bar{μ}_c^N - \bar{μ}_c^F \right) / e.
\end{equation}

We assume a fixed current $J_0 = \abs{J_0} > 0$
flows down through the contact at $x = 0$
and to the left in the semiconductor for $x ≤ 0$,
and no current flows for $x > 0$.
The experimentally measured quantity is the
nonlocal resistance $\rNL = V_L / I_0$,
where $I_0 = - W L J_0 / e$ is the current through the contact at $x = 0$.
It is convenient to introduce the effective nonlocal resistance $\rNLeff$
defined by
\begin{equation}
  \rNLeff = W L \rNL = - e V_L / J_0 = \frac{\bar{μ}_c^F - \bar{μ}_c^N}{J_0} .
\end{equation}
To determine $\rNL$, we must express the difference
of these chemical potentials in terms of $μ_s^N (L)$.

Since there are two ferromagnetic contacts,
we have separate functions $ψ$ and $φ$ for each contact
which we will denote by $ψ^0$, $φ^0$, and $ψ^L$, $φ^L$.
From \cref{eq:diffusion.ferromagnet.solution}, we have
\begin{subequations}
  \begin{align}
    φ^0 \left( z' \right) & = φ_0 e^{k_F z'} , \\
    φ^L \left( z' \right) & = φ_L e^{k_F z'} .
  \end{align}
\end{subequations}

The physical restriction on the current flow in the semiconductor
is imposed by noting that since $σ_s^N = 0$,
\cref{eq:currents.current} gives $J^N = σ^N ∇v$, so we must have
\begin{equation}
  v_x (x) =
    \begin{cases}
      v_x (0) - \left( J_0 / σ^N \right) x & \text{ for } x ≤ 0 , \\
      v_x (0)                              & \text{ for } x > 0 ,
    \end{cases}
\end{equation}
$v_y (x) = v_y (0)$, and $v_z (x) = v_z (0)$.

Using \cref{eq:currents.current},
the restriction on the current flow in each ferromagnet gives
\begin{subequations}
  \begin{align}
    ∇ψ^0 & = \left( J_0 / σ^F \right) - P_σ^F ∇φ^0 , \\
    ∇ψ^L & = - P_σ^F ∇φ^L .
  \end{align}
\end{subequations}
Integrating and enforcing
$e V_c = v_x (0) - \left( ψ_c - P_σ^F φ_c \right)$,
\begin{subequations}
  \begin{align}
    & \begin{aligned}
        ψ^0 \left( z' \right) & = - e V_0 + P_σ^F φ_0 \left( 2 - e^{k_F z'} \right) \\
                              & \qquad + v_x (0) + \left( J_0 / σ^F \right) z' ,
      \end{aligned} \\
    & \begin{aligned}
        ψ^L \left( z' \right) & = - e V_L + P_σ^F φ_L \left( 2 - e^{k_F z'} \right) \\
                              & \qquad + v_x (0) .
      \end{aligned}
  \end{align}
\end{subequations}

There is no current at the contact at $x = 0$,
thus \cref{eq:contact_currents.2.current} gives
\begin{equation}
  ψ_L - v_L = P_Σ^L \left( u_L - φ_L \right) ,
\end{equation}
and with
\cref{eq:bdry_solutions.zero.potential},
we find
\begin{equation}
  \label{eq:rnl.full}
  \begin{aligned}
    \rNLeff & = \left( ψ_L - v_L \right) - P_σ^F φ_L \\
            & = \left[ P_Σ^L \left( 1 - \frac{R_F}{R_F + R_C^L} \right) - \frac{P_σ^F R_F}{R_F + R_C^L} \right] \frac{u_x (L)}{J_0} .
  \end{aligned}
\end{equation}

\subsection{Diffusion equation}

In this section we show how to solve for $μ_s^N (L)$.
This method is based on the one described in
\cite{PhysRevB.80.214427}.
Inside the semiconductor, $u$ satisfies the diffusion equation
\begin{equation}
  \label{eq:diffusion}
  D ∇^2 u - \frac{u}{τ} + ω × u = 0 .
\end{equation}
Here, $D$ is the diffusion constant, $τ$ the spin lifetime,
and $ω = \left( g μ_B / ℏ \right) B$ is proportional to the applied magnetic field
(with $g$ the gyromagnetic ratio and $μ_B$ the Bohr magneton).
The spin diffusion length in the semiconductor is $λ = 1 / k = \sqrt{D τ}$.

The function $u = u(x)$ only varies along $x$,
and we introduce the notation
\begin{equation}
  u_x (x) =
    \begin{cases}
      u_{x-} (x) \text{ for } x < 0     , \\
      u_{x0} (x) \text{ for } 0 ≤ x ≤ L , \\
      u_{x+} (x) \text{ for }     L < x ,
    \end{cases}
\end{equation}
with similar expressions for $u_y$ and $u_z$.
The most general solution to \cref{eq:diffusion} decouples $u_z$ from $u_x$ and $u_y$.
The requirement $\lim_{x → {±} ∞} u(x) = 0$ yields
\begin{subequations}
  \begin{alignat}{3}
    & u_{z{±}} (x) && {}={} && A^∓ e^{∓ k x}                     , \\
    & u_{z0} (x) && {}={} && A_0^+ e^{k x} {}+{} A_0^- e^{-k x} ,
  \end{alignat}
\end{subequations}
and
\begin{subequations}
  \begin{alignat}{6}
    & u_{x{±}} (x) && =   && B^∓ e^{∓ κ x} && {}+{} &&   && C^∓ e^{∓ \bar{κ} x} , \\
    & u_{y{±}} (x) && = i && B^∓ e^{∓ κ x} && {}-{} && i && C^∓ e^{∓ \bar{κ} x} ,
  \end{alignat}
  \begin{alignat}{12}
    & u_{x0} (x) && =   && B_0^+ e^{κ x} && {}+{} &&   && B_0^- e^{- κ x} && {}+{} &&   && C_0^+ e^{ \bar{κ} x } && {}+{} &&   && C_0^- e^{ - \bar{κ} x } , \\
    & u_{y0} (x) && = i && B_0^+ e^{κ x} && {}+{} && i && B_0^- e^{- κ x} && {}-{} && i && C_0^+ e^{ \bar{κ} x } && {}-{} && i && C_0^- e^{ - \bar{κ} x } ,
  \end{alignat}
\end{subequations}
where $κ = k \sqrt{1 + i ω τ}$.
The twelve constants $A$, $B$ and $C$ (with their various subscripts and superscripts) must be determined by imposing the appropriate boundary conditions.

We first require $u$ be continuous at $x = 0$ and $x = L$; this gives six equations.
We now require a boundary condition on $∇u$, but $∇u$ cannot be assumed continuous at the contact.
We make the assumption that the total spin current at the contact
is the sum of the spin currents on either side, i.e.,
\begin{subequations}
  \label{eq:current.sum}
  \begin{alignat}{2}
    & \jmath_0 && = σ^N \left[ - u_-'(0) + u_0'(0) \right] , \\
    & \jmath_L && = σ^N \left[ - u_0'(L) + u_+'(L) \right] .
  \end{alignat}
\end{subequations}
The signs have been chosen to be consistent with the physical geometry.
The only nonzero component of the current at the contacts inside the semiconductor
is the $x$ component at $x = 0$, so we use \cref{eq:bdry_solutions.current}.
For all other components there is zero current at the contact, and we use
\cref{eq:bdry_solutions.zero.current}.
Together with \cref{eq:current.sum}, this gives the other six equations,
\begin{subequations}
  \begin{alignat}{8}
    & - & {} & u_{z-}'(0) & {}+{} & u_{z0}'(0) & {}+{} & η_0 u_z(0) && = 0 , \\
    &   & {} & u_{z+}'(L) & {}-{} & u_{z0}'(L) & {}+{} & η_L u_z(L) && = 0 , \\
    & - & {} & u_{x-}'(0) & {}+{} & u_{x0}'(0) & {}+{} & η_0 u_x(0) && = Δ , \\
    &   & {} & u_{x+}'(L) & {}-{} & u_{x0}'(L) & {}+{} & η_L u_x(L) && = 0 , \\
    & - & {} & u_{y-}'(0) & {}+{} & u_{y0}'(0) & {}+{} & η_0 u_y(0) && = 0 , \\
    &   & {} & u_{y+}'(L) & {}-{} & u_{y0}'(L) & {}+{} & η_L u_y(L) && = 0 ,
  \end{alignat}
\end{subequations}
where
\begin{subequations}
  \begin{alignat}{2}
  & η_i^{-1} && = - σ^N \left( R_F + R_C^i \right) , \\
  & Δ && = - (- J_0) \left( P_σ^F R_F + P_Σ^0 R_C^0 \right) η_0 .
  \end{alignat}
\end{subequations}
We define the $r$-parameter, $r_i = - η_i^{-1}$, introduced in \cref{eq:r-parameter}.

These equations can be organized into a matrix equation and solved algebraically.
A solution for $u_z$ corresponds to a condition of vanishing determinant,
\begin{equation}
  e^{-2 L / λ} = \left( 1 + \frac{2 r_0}{λ} \right) \left( 1 + \frac{2 r_L}{λ} \right) ,
\end{equation}
which can never be satisfied
\footnote{
  Except at the nonphysical point $L / λ = r_i / λ = 0$.
},
thus $u_z = 0$ is the only allowed solution.
The other two components form an eight dimensional linear system.
Solving this gives the remaining constants, and thus
$u_x (L) = e^{- κ L} B^- + e^{- \bar{κ} L} C^-$.

Finally, by using $p_1 = - σ^N Δ / J_0$ along with \cref{eq:rnl.full},
we can introduce $\rSQ$ from \cref{eq:square_resistance}
and the polarizations
\begin{subequations}
  \label{eq:polarizations}
  \begin{align}
    p_1 & = \frac{P_σ^F R_F + P_Σ^L R_C^L}{R_F + R_C^L} , \\
    p_2 / p_1 & = \left. \left( 1 - \frac{P_σ^F R_F}{P_Σ^L R_C^L} \right) \middle/
                  \left(1 + \frac{P_σ^F R_F}{P_Σ^L R_C^L} \right) \right. ,
  \end{align}
\end{subequations}
to write
\begin{equation}
  \frac{\rNLeff}{\rSQ} = \frac{p_1 p_2}{W / λ} \left[ - \frac{k u_x (L)}{Δ} \right] .
\end{equation}
The factor in brackets is the function $f$ given in \cref{eq:f}.


\begin{thebibliography}{34}%
\makeatletter
\providecommand \@ifxundefined [1]{%
 \@ifx{#1\undefined}
}%
\providecommand \@ifnum [1]{%
 \ifnum #1\expandafter \@firstoftwo
 \else \expandafter \@secondoftwo
 \fi
}%
\providecommand \@ifx [1]{%
 \ifx #1\expandafter \@firstoftwo
 \else \expandafter \@secondoftwo
 \fi
}%
\providecommand \natexlab [1]{#1}%
\providecommand \enquote  [1]{``#1''}%
\providecommand \bibnamefont  [1]{#1}%
\providecommand \bibfnamefont [1]{#1}%
\providecommand \citenamefont [1]{#1}%
\providecommand \href@noop [0]{\@secondoftwo}%
\providecommand \href [0]{\begingroup \@sanitize@url \@href}%
\providecommand \@href[1]{\@@startlink{#1}\@@href}%
\providecommand \@@href[1]{\endgroup#1\@@endlink}%
\providecommand \@sanitize@url [0]{\catcode `\\12\catcode `\$12\catcode
  `\&12\catcode `\#12\catcode `\^12\catcode `\_12\catcode `\%12\relax}%
\providecommand \@@startlink[1]{}%
\providecommand \@@endlink[0]{}%
\providecommand \url  [0]{\begingroup\@sanitize@url \@url }%
\providecommand \@url [1]{\endgroup\@href {#1}{\urlprefix }}%
\providecommand \urlprefix  [0]{URL }%
\providecommand \Eprint [0]{\href }%
\providecommand \doibase [0]{http://dx.doi.org/}%
\providecommand \selectlanguage [0]{\@gobble}%
\providecommand \bibinfo  [0]{\@secondoftwo}%
\providecommand \bibfield  [0]{\@secondoftwo}%
\providecommand \translation [1]{[#1]}%
\providecommand \BibitemOpen [0]{}%
\providecommand \bibitemStop [0]{}%
\providecommand \bibitemNoStop [0]{.\EOS\space}%
\providecommand \EOS [0]{\spacefactor3000\relax}%
\providecommand \BibitemShut  [1]{\csname bibitem#1\endcsname}%
\let\auto@bib@innerbib\@empty
\bibitem [{\citenamefont {Wolf}\ \emph {et~al.}(2001)\citenamefont {Wolf},
  \citenamefont {Awschalom}, \citenamefont {Buhrman}, \citenamefont {Daughton},
  \citenamefont {von Molnár}, \citenamefont {Roukes}, \citenamefont
  {Chtchelkanova},\ and\ \citenamefont {Treger}}]{Wolf16112001}%
  \BibitemOpen
  \bibfield  {author} {\bibinfo {author} {\bibfnamefont {S.~A.}\ \bibnamefont
  {Wolf}}, \bibinfo {author} {\bibfnamefont {D.~D.}\ \bibnamefont {Awschalom}},
  \bibinfo {author} {\bibfnamefont {R.~A.}\ \bibnamefont {Buhrman}}, \bibinfo
  {author} {\bibfnamefont {J.~M.}\ \bibnamefont {Daughton}}, \bibinfo {author}
  {\bibfnamefont {S.}~\bibnamefont {von Molnár}}, \bibinfo {author}
  {\bibfnamefont {M.~L.}\ \bibnamefont {Roukes}}, \bibinfo {author}
  {\bibfnamefont {A.~Y.}\ \bibnamefont {Chtchelkanova}}, \ and\ \bibinfo
  {author} {\bibfnamefont {D.~M.}\ \bibnamefont {Treger}},\ }\href {\doibase
  10.1126/science.1065389} {\bibfield  {journal} {\bibinfo  {journal}
  {Science}\ }\textbf {\bibinfo {volume} {294}},\ \bibinfo {pages} {1488}
  (\bibinfo {year} {2001})}\BibitemShut {NoStop}%
\bibitem [{\citenamefont {\v{Z}uti\'{c}}\ \emph {et~al.}(2004)\citenamefont
  {\v{Z}uti\'{c}}, \citenamefont {Fabian},\ and\ \citenamefont
  {Das~Sarma}}]{RevModPhys.76.323}%
  \BibitemOpen
  \bibfield  {author} {\bibinfo {author} {\bibfnamefont {I.}~\bibnamefont
  {\v{Z}uti\'{c}}}, \bibinfo {author} {\bibfnamefont {J.}~\bibnamefont
  {Fabian}}, \ and\ \bibinfo {author} {\bibfnamefont {S.}~\bibnamefont
  {Das~Sarma}},\ }\href {\doibase 10.1103/RevModPhys.76.323} {\bibfield
  {journal} {\bibinfo  {journal} {Rev. Mod. Phys.}\ }\textbf {\bibinfo {volume}
  {76}},\ \bibinfo {pages} {323} (\bibinfo {year} {2004})}\BibitemShut
  {NoStop}%
\bibitem [{\citenamefont {{Datta}}\ and\ \citenamefont
  {{Das}}(1990)}]{1990ApPhL..56..665D}%
  \BibitemOpen
  \bibfield  {author} {\bibinfo {author} {\bibfnamefont {S.}~\bibnamefont
  {{Datta}}}\ and\ \bibinfo {author} {\bibfnamefont {B.}~\bibnamefont
  {{Das}}},\ }\href {\doibase 10.1063/1.102730} {\bibfield  {journal} {\bibinfo
   {journal} {Applied Physics Letters}\ }\textbf {\bibinfo {volume} {56}},\
  \bibinfo {pages} {665} (\bibinfo {year} {1990})}\BibitemShut {NoStop}%
\bibitem [{\citenamefont {Jedema}\ \emph {et~al.}(2001)\citenamefont {Jedema},
  \citenamefont {Filip},\ and\ \citenamefont {van Wees}}]{Jedema2001}%
  \BibitemOpen
  \bibfield  {author} {\bibinfo {author} {\bibfnamefont {F.~J.}\ \bibnamefont
  {Jedema}}, \bibinfo {author} {\bibfnamefont {A.~T.}\ \bibnamefont {Filip}}, \
  and\ \bibinfo {author} {\bibfnamefont {B.~J.}\ \bibnamefont {van Wees}},\
  }\href {\doibase 10.1038/35066533} {\bibfield  {journal} {\bibinfo  {journal}
  {Nature}\ }\textbf {\bibinfo {volume} {410}},\ \bibinfo {pages} {345}
  (\bibinfo {year} {2001})}\BibitemShut {NoStop}%
\bibitem [{\citenamefont {Johnson}\ and\ \citenamefont
  {Silsbee}(1985)}]{PhysRevLett.55.1790}%
  \BibitemOpen
  \bibfield  {author} {\bibinfo {author} {\bibfnamefont {M.}~\bibnamefont
  {Johnson}}\ and\ \bibinfo {author} {\bibfnamefont {R.~H.}\ \bibnamefont
  {Silsbee}},\ }\href {\doibase 10.1103/PhysRevLett.55.1790} {\bibfield
  {journal} {\bibinfo  {journal} {Phys. Rev. Lett.}\ }\textbf {\bibinfo
  {volume} {55}},\ \bibinfo {pages} {1790} (\bibinfo {year}
  {1985})}\BibitemShut {NoStop}%
\bibitem [{\citenamefont {Yang}\ \emph {et~al.}(2008)\citenamefont {Yang},
  \citenamefont {Kimura},\ and\ \citenamefont {Otani}}]{Yang2008}%
  \BibitemOpen
  \bibfield  {author} {\bibinfo {author} {\bibfnamefont {T.}~\bibnamefont
  {Yang}}, \bibinfo {author} {\bibfnamefont {T.}~\bibnamefont {Kimura}}, \ and\
  \bibinfo {author} {\bibfnamefont {Y.}~\bibnamefont {Otani}},\ }\href
  {\doibase 10.1038/nphys1095} {\bibfield  {journal} {\bibinfo  {journal} {Nat
  Phys}\ }\textbf {\bibinfo {volume} {4}},\ \bibinfo {pages} {851} (\bibinfo
  {year} {2008})}\BibitemShut {NoStop}%
\bibitem [{\citenamefont {Valenzuela}\ \emph {et~al.}(2005)\citenamefont
  {Valenzuela}, \citenamefont {Monsma}, \citenamefont {Marcus}, \citenamefont
  {Narayanamurti},\ and\ \citenamefont {Tinkham}}]{PhysRevLett.94.196601}%
  \BibitemOpen
  \bibfield  {author} {\bibinfo {author} {\bibfnamefont {S.~O.}\ \bibnamefont
  {Valenzuela}}, \bibinfo {author} {\bibfnamefont {D.~J.}\ \bibnamefont
  {Monsma}}, \bibinfo {author} {\bibfnamefont {C.~M.}\ \bibnamefont {Marcus}},
  \bibinfo {author} {\bibfnamefont {V.}~\bibnamefont {Narayanamurti}}, \ and\
  \bibinfo {author} {\bibfnamefont {M.}~\bibnamefont {Tinkham}},\ }\href
  {\doibase 10.1103/PhysRevLett.94.196601} {\bibfield  {journal} {\bibinfo
  {journal} {Phys. Rev. Lett.}\ }\textbf {\bibinfo {volume} {94}},\ \bibinfo
  {pages} {196601} (\bibinfo {year} {2005})}\BibitemShut {NoStop}%
\bibitem [{\citenamefont {Novoselov}\ \emph {et~al.}(2004)\citenamefont
  {Novoselov}, \citenamefont {Geim}, \citenamefont {Morozov}, \citenamefont
  {Jiang}, \citenamefont {Zhang}, \citenamefont {Dubonos}, \citenamefont
  {Grigorieva},\ and\ \citenamefont {Firsov}}]{Novoselov22102004}%
  \BibitemOpen
  \bibfield  {author} {\bibinfo {author} {\bibfnamefont {K.~S.}\ \bibnamefont
  {Novoselov}}, \bibinfo {author} {\bibfnamefont {A.~K.}\ \bibnamefont {Geim}},
  \bibinfo {author} {\bibfnamefont {S.~V.}\ \bibnamefont {Morozov}}, \bibinfo
  {author} {\bibfnamefont {D.}~\bibnamefont {Jiang}}, \bibinfo {author}
  {\bibfnamefont {Y.}~\bibnamefont {Zhang}}, \bibinfo {author} {\bibfnamefont
  {S.~V.}\ \bibnamefont {Dubonos}}, \bibinfo {author} {\bibfnamefont {I.~V.}\
  \bibnamefont {Grigorieva}}, \ and\ \bibinfo {author} {\bibfnamefont {A.~A.}\
  \bibnamefont {Firsov}},\ }\href {\doibase 10.1126/science.1102896} {\bibfield
   {journal} {\bibinfo  {journal} {Science}\ }\textbf {\bibinfo {volume}
  {306}},\ \bibinfo {pages} {666} (\bibinfo {year} {2004})}\BibitemShut
  {NoStop}%
\bibitem [{\citenamefont {Huertas-Hernando}\ \emph {et~al.}(2006)\citenamefont
  {Huertas-Hernando}, \citenamefont {Guinea},\ and\ \citenamefont
  {Brataas}}]{PhysRevB.74.155426}%
  \BibitemOpen
  \bibfield  {author} {\bibinfo {author} {\bibfnamefont {D.}~\bibnamefont
  {Huertas-Hernando}}, \bibinfo {author} {\bibfnamefont {F.}~\bibnamefont
  {Guinea}}, \ and\ \bibinfo {author} {\bibfnamefont {A.}~\bibnamefont
  {Brataas}},\ }\href {\doibase 10.1103/PhysRevB.74.155426} {\bibfield
  {journal} {\bibinfo  {journal} {Phys. Rev. B}\ }\textbf {\bibinfo {volume}
  {74}},\ \bibinfo {pages} {155426} (\bibinfo {year} {2006})}\BibitemShut
  {NoStop}%
\bibitem [{\citenamefont {Trauzettel}\ \emph {et~al.}(2007)\citenamefont
  {Trauzettel}, \citenamefont {Bulaev}, \citenamefont {Loss},\ and\
  \citenamefont {Burkard}}]{Trauzettel2007}%
  \BibitemOpen
  \bibfield  {author} {\bibinfo {author} {\bibfnamefont {B.}~\bibnamefont
  {Trauzettel}}, \bibinfo {author} {\bibfnamefont {D.~V.}\ \bibnamefont
  {Bulaev}}, \bibinfo {author} {\bibfnamefont {D.}~\bibnamefont {Loss}}, \ and\
  \bibinfo {author} {\bibfnamefont {G.}~\bibnamefont {Burkard}},\ }\href
  {\doibase 10.1038/nphys544} {\bibfield  {journal} {\bibinfo  {journal} {Nat
  Phys}\ }\textbf {\bibinfo {volume} {3}},\ \bibinfo {pages} {192} (\bibinfo
  {year} {2007})}\BibitemShut {NoStop}%
\bibitem [{\citenamefont {Tombros}\ \emph {et~al.}(2007)\citenamefont
  {Tombros}, \citenamefont {Jozsa}, \citenamefont {Popinciuc}, \citenamefont
  {Jonkman},\ and\ \citenamefont {van Wees}}]{Tombros2007}%
  \BibitemOpen
  \bibfield  {author} {\bibinfo {author} {\bibfnamefont {N.}~\bibnamefont
  {Tombros}}, \bibinfo {author} {\bibfnamefont {C.}~\bibnamefont {Jozsa}},
  \bibinfo {author} {\bibfnamefont {M.}~\bibnamefont {Popinciuc}}, \bibinfo
  {author} {\bibfnamefont {H.~T.}\ \bibnamefont {Jonkman}}, \ and\ \bibinfo
  {author} {\bibfnamefont {B.~J.}\ \bibnamefont {van Wees}},\ }\href {\doibase
  10.1038/nature06037} {\bibfield  {journal} {\bibinfo  {journal} {Nature}\
  }\textbf {\bibinfo {volume} {448}},\ \bibinfo {pages} {571} (\bibinfo {year}
  {2007})}\BibitemShut {NoStop}%
\bibitem [{\citenamefont {Ohishi}\ \emph {et~al.}(2007)\citenamefont {Ohishi},
  \citenamefont {Shiraishi}, \citenamefont {Nouchi}, \citenamefont {Nozaki},
  \citenamefont {Shinjo},\ and\ \citenamefont {Suzuki}}]{JJAP.46.L605}%
  \BibitemOpen
  \bibfield  {author} {\bibinfo {author} {\bibfnamefont {M.}~\bibnamefont
  {Ohishi}}, \bibinfo {author} {\bibfnamefont {M.}~\bibnamefont {Shiraishi}},
  \bibinfo {author} {\bibfnamefont {R.}~\bibnamefont {Nouchi}}, \bibinfo
  {author} {\bibfnamefont {T.}~\bibnamefont {Nozaki}}, \bibinfo {author}
  {\bibfnamefont {T.}~\bibnamefont {Shinjo}}, \ and\ \bibinfo {author}
  {\bibfnamefont {Y.}~\bibnamefont {Suzuki}},\ }\href {\doibase
  10.7567/JJAP.46.L605} {\bibfield  {journal} {\bibinfo  {journal} {Japanese
  Journal of Applied Physics}\ }\textbf {\bibinfo {volume} {46}},\ \bibinfo
  {pages} {L605} (\bibinfo {year} {2007})}\BibitemShut {NoStop}%
\bibitem [{\citenamefont {Cho}\ \emph {et~al.}(2007)\citenamefont {Cho},
  \citenamefont {Chen},\ and\ \citenamefont {Fuhrer}}]{Cho2007}%
  \BibitemOpen
  \bibfield  {author} {\bibinfo {author} {\bibfnamefont {S.}~\bibnamefont
  {Cho}}, \bibinfo {author} {\bibfnamefont {Y.-F.}\ \bibnamefont {Chen}}, \
  and\ \bibinfo {author} {\bibfnamefont {M.~S.}\ \bibnamefont {Fuhrer}},\
  }\href {\doibase http://dx.doi.org/10.1063/1.2784934} {\bibfield  {journal}
  {\bibinfo  {journal} {Applied Physics Letters}\ }\textbf {\bibinfo {volume}
  {91}},\ \bibinfo {eid} {123105} (\bibinfo {year} {2007})}\BibitemShut
  {NoStop}%
\bibitem [{\citenamefont {Tombros}\ \emph {et~al.}(2008)\citenamefont
  {Tombros}, \citenamefont {Tanabe}, \citenamefont {Veligura}, \citenamefont
  {Jozsa}, \citenamefont {Popinciuc}, \citenamefont {Jonkman},\ and\
  \citenamefont {van Wees}}]{PhysRevLett.101.046601}%
  \BibitemOpen
  \bibfield  {author} {\bibinfo {author} {\bibfnamefont {N.}~\bibnamefont
  {Tombros}}, \bibinfo {author} {\bibfnamefont {S.}~\bibnamefont {Tanabe}},
  \bibinfo {author} {\bibfnamefont {A.}~\bibnamefont {Veligura}}, \bibinfo
  {author} {\bibfnamefont {C.}~\bibnamefont {Jozsa}}, \bibinfo {author}
  {\bibfnamefont {M.}~\bibnamefont {Popinciuc}}, \bibinfo {author}
  {\bibfnamefont {H.~T.}\ \bibnamefont {Jonkman}}, \ and\ \bibinfo {author}
  {\bibfnamefont {B.~J.}\ \bibnamefont {van Wees}},\ }\href {\doibase
  10.1103/PhysRevLett.101.046601} {\bibfield  {journal} {\bibinfo  {journal}
  {Phys. Rev. Lett.}\ }\textbf {\bibinfo {volume} {101}},\ \bibinfo {pages}
  {046601} (\bibinfo {year} {2008})}\BibitemShut {NoStop}%
\bibitem [{\citenamefont {Hill}\ \emph {et~al.}(2006)\citenamefont {Hill},
  \citenamefont {Geim}, \citenamefont {Novoselov}, \citenamefont {Schedin},\
  and\ \citenamefont {Blake}}]{1704408}%
  \BibitemOpen
  \bibfield  {author} {\bibinfo {author} {\bibfnamefont {E.}~\bibnamefont
  {Hill}}, \bibinfo {author} {\bibfnamefont {A.}~\bibnamefont {Geim}}, \bibinfo
  {author} {\bibfnamefont {K.}~\bibnamefont {Novoselov}}, \bibinfo {author}
  {\bibfnamefont {F.}~\bibnamefont {Schedin}}, \ and\ \bibinfo {author}
  {\bibfnamefont {P.}~\bibnamefont {Blake}},\ }\href {\doibase
  10.1109/TMAG.2006.878852} {\bibfield  {journal} {\bibinfo  {journal}
  {Magnetics, IEEE Transactions on}\ }\textbf {\bibinfo {volume} {42}},\
  \bibinfo {pages} {2694} (\bibinfo {year} {2006})}\BibitemShut {NoStop}%
\bibitem [{\citenamefont {Han}\ \emph {et~al.}(2012{\natexlab{a}})\citenamefont
  {Han}, \citenamefont {Chen}, \citenamefont {Wang}, \citenamefont {McCreary},
  \citenamefont {Wen}, \citenamefont {Swartz}, \citenamefont {Shi},\ and\
  \citenamefont {Kawakami}}]{Han2012}%
  \BibitemOpen
  \bibfield  {author} {\bibinfo {author} {\bibfnamefont {W.}~\bibnamefont
  {Han}}, \bibinfo {author} {\bibfnamefont {J.-R.}\ \bibnamefont {Chen}},
  \bibinfo {author} {\bibfnamefont {D.}~\bibnamefont {Wang}}, \bibinfo {author}
  {\bibfnamefont {K.~M.}\ \bibnamefont {McCreary}}, \bibinfo {author}
  {\bibfnamefont {H.}~\bibnamefont {Wen}}, \bibinfo {author} {\bibfnamefont
  {A.~G.}\ \bibnamefont {Swartz}}, \bibinfo {author} {\bibfnamefont
  {J.}~\bibnamefont {Shi}}, \ and\ \bibinfo {author} {\bibfnamefont {R.~K.}\
  \bibnamefont {Kawakami}},\ }\href {\doibase 10.1021/nl301567n} {\bibfield
  {journal} {\bibinfo  {journal} {Nano Letters}\ }\textbf {\bibinfo {volume}
  {12}},\ \bibinfo {pages} {3443–3447} (\bibinfo {year}
  {2012}{\natexlab{a}})}\BibitemShut {NoStop}%
\bibitem [{\citenamefont {Han}\ \emph {et~al.}(2012{\natexlab{b}})\citenamefont
  {Han}, \citenamefont {McCreary}, \citenamefont {Pi}, \citenamefont {Wang},
  \citenamefont {Li}, \citenamefont {Wen}, \citenamefont {Chen},\ and\
  \citenamefont {Kawakami}}]{Han2012369}%
  \BibitemOpen
  \bibfield  {author} {\bibinfo {author} {\bibfnamefont {W.}~\bibnamefont
  {Han}}, \bibinfo {author} {\bibfnamefont {K.}~\bibnamefont {McCreary}},
  \bibinfo {author} {\bibfnamefont {K.}~\bibnamefont {Pi}}, \bibinfo {author}
  {\bibfnamefont {W.}~\bibnamefont {Wang}}, \bibinfo {author} {\bibfnamefont
  {Y.}~\bibnamefont {Li}}, \bibinfo {author} {\bibfnamefont {H.}~\bibnamefont
  {Wen}}, \bibinfo {author} {\bibfnamefont {J.}~\bibnamefont {Chen}}, \ and\
  \bibinfo {author} {\bibfnamefont {R.}~\bibnamefont {Kawakami}},\ }\href
  {\doibase 10.1016/j.jmmm.2011.08.001} {\bibfield  {journal} {\bibinfo
  {journal} {Journal of Magnetism and Magnetic Materials}\ }\textbf {\bibinfo
  {volume} {324}},\ \bibinfo {pages} {369} (\bibinfo {year}
  {2012}{\natexlab{b}})}\BibitemShut {NoStop}%
\bibitem [{\citenamefont {Józsa}\ \emph {et~al.}(2009)\citenamefont {Józsa},
  \citenamefont {Maassen}, \citenamefont {Popinciuc}, \citenamefont {Zomer},
  \citenamefont {Veligura}, \citenamefont {Jonkman},\ and\ \citenamefont {van
  Wees}}]{PhysRevB.80.241403}%
  \BibitemOpen
  \bibfield  {author} {\bibinfo {author} {\bibfnamefont {C.}~\bibnamefont
  {Józsa}}, \bibinfo {author} {\bibfnamefont {T.}~\bibnamefont {Maassen}},
  \bibinfo {author} {\bibfnamefont {M.}~\bibnamefont {Popinciuc}}, \bibinfo
  {author} {\bibfnamefont {P.~J.}\ \bibnamefont {Zomer}}, \bibinfo {author}
  {\bibfnamefont {A.}~\bibnamefont {Veligura}}, \bibinfo {author}
  {\bibfnamefont {H.~T.}\ \bibnamefont {Jonkman}}, \ and\ \bibinfo {author}
  {\bibfnamefont {B.~J.}\ \bibnamefont {van Wees}},\ }\href {\doibase
  10.1103/PhysRevB.80.241403} {\bibfield  {journal} {\bibinfo  {journal} {Phys.
  Rev. B}\ }\textbf {\bibinfo {volume} {80}},\ \bibinfo {pages} {241403}
  (\bibinfo {year} {2009})}\BibitemShut {NoStop}%
\bibitem [{\citenamefont {Popinciuc}\ \emph {et~al.}(2009)\citenamefont
  {Popinciuc}, \citenamefont {Józsa}, \citenamefont {Zomer}, \citenamefont
  {Tombros}, \citenamefont {Veligura}, \citenamefont {Jonkman},\ and\
  \citenamefont {van Wees}}]{PhysRevB.80.214427}%
  \BibitemOpen
  \bibfield  {author} {\bibinfo {author} {\bibfnamefont {M.}~\bibnamefont
  {Popinciuc}}, \bibinfo {author} {\bibfnamefont {C.}~\bibnamefont {Józsa}},
  \bibinfo {author} {\bibfnamefont {P.~J.}\ \bibnamefont {Zomer}}, \bibinfo
  {author} {\bibfnamefont {N.}~\bibnamefont {Tombros}}, \bibinfo {author}
  {\bibfnamefont {A.}~\bibnamefont {Veligura}}, \bibinfo {author}
  {\bibfnamefont {H.~T.}\ \bibnamefont {Jonkman}}, \ and\ \bibinfo {author}
  {\bibfnamefont {B.~J.}\ \bibnamefont {van Wees}},\ }\href {\doibase
  10.1103/PhysRevB.80.214427} {\bibfield  {journal} {\bibinfo  {journal} {Phys.
  Rev. B}\ }\textbf {\bibinfo {volume} {80}},\ \bibinfo {pages} {214427}
  (\bibinfo {year} {2009})}\BibitemShut {NoStop}%
\bibitem [{\citenamefont {Pi}\ \emph {et~al.}(2010)\citenamefont {Pi},
  \citenamefont {Han}, \citenamefont {McCreary}, \citenamefont {Swartz},
  \citenamefont {Li},\ and\ \citenamefont {Kawakami}}]{PhysRevLett.104.187201}%
  \BibitemOpen
  \bibfield  {author} {\bibinfo {author} {\bibfnamefont {K.}~\bibnamefont
  {Pi}}, \bibinfo {author} {\bibfnamefont {W.}~\bibnamefont {Han}}, \bibinfo
  {author} {\bibfnamefont {K.~M.}\ \bibnamefont {McCreary}}, \bibinfo {author}
  {\bibfnamefont {A.~G.}\ \bibnamefont {Swartz}}, \bibinfo {author}
  {\bibfnamefont {Y.}~\bibnamefont {Li}}, \ and\ \bibinfo {author}
  {\bibfnamefont {R.~K.}\ \bibnamefont {Kawakami}},\ }\href {\doibase
  10.1103/PhysRevLett.104.187201} {\bibfield  {journal} {\bibinfo  {journal}
  {Phys. Rev. Lett.}\ }\textbf {\bibinfo {volume} {104}},\ \bibinfo {pages}
  {187201} (\bibinfo {year} {2010})}\BibitemShut {NoStop}%
\bibitem [{\citenamefont {Elliott}(1954)}]{PhysRev.96.266}%
  \BibitemOpen
  \bibfield  {author} {\bibinfo {author} {\bibfnamefont {R.~J.}\ \bibnamefont
  {Elliott}},\ }\href {\doibase 10.1103/PhysRev.96.266} {\bibfield  {journal}
  {\bibinfo  {journal} {Phys. Rev.}\ }\textbf {\bibinfo {volume} {96}},\
  \bibinfo {pages} {266} (\bibinfo {year} {1954})}\BibitemShut {NoStop}%
\bibitem [{\citenamefont {Han}\ and\ \citenamefont
  {Kawakami}(2011)}]{PhysRevLett.107.047207}%
  \BibitemOpen
  \bibfield  {author} {\bibinfo {author} {\bibfnamefont {W.}~\bibnamefont
  {Han}}\ and\ \bibinfo {author} {\bibfnamefont {R.~K.}\ \bibnamefont
  {Kawakami}},\ }\href {\doibase 10.1103/PhysRevLett.107.047207} {\bibfield
  {journal} {\bibinfo  {journal} {Phys. Rev. Lett.}\ }\textbf {\bibinfo
  {volume} {107}},\ \bibinfo {pages} {047207} (\bibinfo {year}
  {2011})}\BibitemShut {NoStop}%
\bibitem [{\citenamefont {Takahashi}\ and\ \citenamefont
  {Maekawa}(2003)}]{PhysRevB.67.052409}%
  \BibitemOpen
  \bibfield  {author} {\bibinfo {author} {\bibfnamefont {S.}~\bibnamefont
  {Takahashi}}\ and\ \bibinfo {author} {\bibfnamefont {S.}~\bibnamefont
  {Maekawa}},\ }\href {\doibase 10.1103/PhysRevB.67.052409} {\bibfield
  {journal} {\bibinfo  {journal} {Phys. Rev. B}\ }\textbf {\bibinfo {volume}
  {67}},\ \bibinfo {pages} {052409} (\bibinfo {year} {2003})}\BibitemShut
  {NoStop}%
\bibitem [{\citenamefont {Wojtaszek}\ \emph {et~al.}(2014)\citenamefont
  {Wojtaszek}, \citenamefont {Vera-Marun},\ and\ \citenamefont {van
  Wees}}]{1404.6276v1}%
  \BibitemOpen
  \bibfield  {author} {\bibinfo {author} {\bibfnamefont {M.}~\bibnamefont
  {Wojtaszek}}, \bibinfo {author} {\bibfnamefont {I.~J.}\ \bibnamefont
  {Vera-Marun}}, \ and\ \bibinfo {author} {\bibfnamefont {B.~J.}\ \bibnamefont
  {van Wees}},\ }\href {http://arxiv.org/abs/1404.6276v1;
  http://arxiv.org/pdf/1404.6276v1} {\  (\bibinfo {year} {2014})},\ \Eprint
  {http://arxiv.org/abs/1404.6276v1} {arXiv:1404.6276v1 [cond-mat.mes-hall]}
  \BibitemShut {NoStop}%
\bibitem [{Note1()}]{Note1}%
  \BibitemOpen
  \bibinfo {note} {Assuming the polarizations $P_\IeC {\protect \ensuremath
  {\sigma }}^F$ and $P_\IeC {\protect \ensuremath {\Sigma }}^L$ have the same
  sign bounds $P \IeC {\protect \ensuremath {\leq }} 1$.}\BibitemShut {Stop}%
\bibitem [{\citenamefont {Han}\ \emph {et~al.}(2010)\citenamefont {Han},
  \citenamefont {Pi}, \citenamefont {McCreary}, \citenamefont {Li},
  \citenamefont {Wong}, \citenamefont {Swartz},\ and\ \citenamefont
  {Kawakami}}]{PhysRevLett.105.167202}%
  \BibitemOpen
  \bibfield  {author} {\bibinfo {author} {\bibfnamefont {W.}~\bibnamefont
  {Han}}, \bibinfo {author} {\bibfnamefont {K.}~\bibnamefont {Pi}}, \bibinfo
  {author} {\bibfnamefont {K.~M.}\ \bibnamefont {McCreary}}, \bibinfo {author}
  {\bibfnamefont {Y.}~\bibnamefont {Li}}, \bibinfo {author} {\bibfnamefont
  {J.~J.~I.}\ \bibnamefont {Wong}}, \bibinfo {author} {\bibfnamefont {A.~G.}\
  \bibnamefont {Swartz}}, \ and\ \bibinfo {author} {\bibfnamefont {R.~K.}\
  \bibnamefont {Kawakami}},\ }\href {\doibase 10.1103/PhysRevLett.105.167202}
  {\bibfield  {journal} {\bibinfo  {journal} {Phys. Rev. Lett.}\ }\textbf
  {\bibinfo {volume} {105}},\ \bibinfo {pages} {167202} (\bibinfo {year}
  {2010})}\BibitemShut {NoStop}%
\bibitem [{\citenamefont {Hunter}(2007)}]{Hunter:2007}%
  \BibitemOpen
  \bibfield  {author} {\bibinfo {author} {\bibfnamefont {J.~D.}\ \bibnamefont
  {Hunter}},\ }\href {\doibase 10.1109/MCSE.2007.55} {\bibfield  {journal}
  {\bibinfo  {journal} {Computing In Science \& Engineering}\ }\textbf
  {\bibinfo {volume} {9}},\ \bibinfo {pages} {90} (\bibinfo {year}
  {2007})}\BibitemShut {NoStop}%
\bibitem [{Note2()}]{Note2}%
  \BibitemOpen
  \bibinfo {note} {An online portal with links to the code used to prepare this
  work is located at \protect \href
  {http://evansosenko.com/spin-lifetime/}{evansosenko.com/spin-lifetime/}.}\BibitemShut
  {Stop}%
\bibitem [{Note3()}]{Note3}%
  \BibitemOpen
  \bibinfo {note} {Parallel and antiparallel data for this device was only
  available at dissimilar field values, thus $\IeC {\protect \ensuremath
  {\Delta }} R_{\protect \text {NL}}$ could not be fit.}\BibitemShut {Stop}%
\bibitem [{\citenamefont {Johnson}\ and\ \citenamefont
  {Silsbee}(1988)}]{PhysRevB.37.5312}%
  \BibitemOpen
  \bibfield  {author} {\bibinfo {author} {\bibfnamefont {M.}~\bibnamefont
  {Johnson}}\ and\ \bibinfo {author} {\bibfnamefont {R.~H.}\ \bibnamefont
  {Silsbee}},\ }\href {\doibase 10.1103/PhysRevB.37.5312} {\bibfield  {journal}
  {\bibinfo  {journal} {Phys. Rev. B}\ }\textbf {\bibinfo {volume} {37}},\
  \bibinfo {pages} {5312} (\bibinfo {year} {1988})}\BibitemShut {NoStop}%
\bibitem [{\citenamefont {Swartz}\ \emph {et~al.}(2013)\citenamefont {Swartz},
  \citenamefont {McCreary}, \citenamefont {Han}, \citenamefont {Wen},\ and\
  \citenamefont {Kawakami}}]{Swartz2013}%
  \BibitemOpen
  \bibfield  {author} {\bibinfo {author} {\bibfnamefont {A.~G.}\ \bibnamefont
  {Swartz}}, \bibinfo {author} {\bibfnamefont {K.~M.}\ \bibnamefont
  {McCreary}}, \bibinfo {author} {\bibfnamefont {W.}~\bibnamefont {Han}},
  \bibinfo {author} {\bibfnamefont {H.}~\bibnamefont {Wen}}, \ and\ \bibinfo
  {author} {\bibfnamefont {R.~K.}\ \bibnamefont {Kawakami}},\ }\href {\doibase
  10.1117/12.2022782} {\enquote {\bibinfo {title} {A systematic approach to
  interpreting hanle spin precession data in non-local spin valves},}\ }
  (\bibinfo {year} {2013})\BibitemShut {NoStop}%
\bibitem [{\citenamefont {Maassen}\ \emph {et~al.}(2012)\citenamefont
  {Maassen}, \citenamefont {Vera-Marun}, \citenamefont {Guimarães},\ and\
  \citenamefont {van Wees}}]{PhysRevB.86.235408}%
  \BibitemOpen
  \bibfield  {author} {\bibinfo {author} {\bibfnamefont {T.}~\bibnamefont
  {Maassen}}, \bibinfo {author} {\bibfnamefont {I.~J.}\ \bibnamefont
  {Vera-Marun}}, \bibinfo {author} {\bibfnamefont {M.~H.~D.}\ \bibnamefont
  {Guimarães}}, \ and\ \bibinfo {author} {\bibfnamefont {B.~J.}\ \bibnamefont
  {van Wees}},\ }\href {\doibase 10.1103/PhysRevB.86.235408} {\bibfield
  {journal} {\bibinfo  {journal} {Phys. Rev. B}\ }\textbf {\bibinfo {volume}
  {86}},\ \bibinfo {pages} {235408} (\bibinfo {year} {2012})}\BibitemShut
  {NoStop}%
\bibitem [{\citenamefont {Fabian}\ \emph {et~al.}(2007)\citenamefont {Fabian},
  \citenamefont {Matos-Abiague}, \citenamefont {Ertler}, \citenamefont
  {Stano},\ and\ \citenamefont
  {\v{Z}uti\'{c}}}]{ActaPhysicaSlovaca.57.4_5.565-907}%
  \BibitemOpen
  \bibfield  {author} {\bibinfo {author} {\bibfnamefont {J.}~\bibnamefont
  {Fabian}}, \bibinfo {author} {\bibfnamefont {A.}~\bibnamefont
  {Matos-Abiague}}, \bibinfo {author} {\bibfnamefont {C.}~\bibnamefont
  {Ertler}}, \bibinfo {author} {\bibfnamefont {P.}~\bibnamefont {Stano}}, \
  and\ \bibinfo {author} {\bibfnamefont {I.}~\bibnamefont {\v{Z}uti\'{c}}},\
  }\href {http://www.physics.sk/aps/pub.php?y=2007&pub=aps-07-04} {\bibfield
  {journal} {\bibinfo  {journal} {Acta Physica Slovaca}\ }\textbf {\bibinfo
  {volume} {57}},\ \bibinfo {pages} {565} (\bibinfo {year} {2007})}\BibitemShut
  {NoStop}%
\bibitem [{Note4()}]{Note4}%
  \BibitemOpen
  \bibinfo {note} {Except at the nonphysical point $L / \IeC {\protect
  \ensuremath {\lambda }} = r_i / \IeC {\protect \ensuremath {\lambda }} =
  0$.}\BibitemShut {Stop}%
\end{thebibliography}
\end{document}